\documentclass{webofc} 
\usepackage[varg]{txfonts}   % Web of Conferences font 

\woctitle{MESON2023}

\def\lamb#1#2{$^{#1}_{\Lambda}${#2}}
\def\lam#1#2{$^{#1}_{~\Lambda}${#2}}

\begin{document} 

\title{Deciphering $\Xi^-$ capture events in light emulsion nuclei} 
\author{\firstname{Eliahu} \lastname{Friedman}\inst{1}\fnsep
\thanks{Eliahu.Friedman@mail.huji.ac.il} \and \firstname{Avraham} 
\lastname{Gal}\inst{1}\fnsep\thanks{avragal@savion.huji.ac.il} 
\institute{Racah Institute of Physics, The Hebrew University, 
Jerusalem 9190401, Israel}} 

\abstract 
{We recently showed that all five KEK and J-PARC uniquely assigned 
two-body $\Xi^- + {{^A}Z} \to {_{\Lambda}^{A^{\prime}}}{Z^{\prime}} + 
{_{\Lambda}^{A^{\prime\prime}}}{Z^{\prime\prime}}$ capture events in CNO 
light emulsion nuclei are consistent with Coulomb-assisted $1p_{\Xi^-}$ 
nuclear states in a $\Xi$-nuclear potential of nuclear-matter depth $V_{\Xi} 
\gtrsim 20$ MeV~\cite{FG21}. Here we argue that the recently reported 
$^{14}$N capture events named KINKA and IRRAWADDY are more likely 
$\Xi^0_{1p}-{^{14}}$C nuclear states~\cite{FG23} than $\Xi^-_{1s}-{^{14}}$N 
states, the latter assignment implying considerably smaller values of 
$V_{\Xi}$.} 
\maketitle

\section{Introduction} 
\label{sec:intro} 

\begin{figure}[!h] 
\centering 
%\sidecaption 
\includegraphics[width=0.6\textwidth]{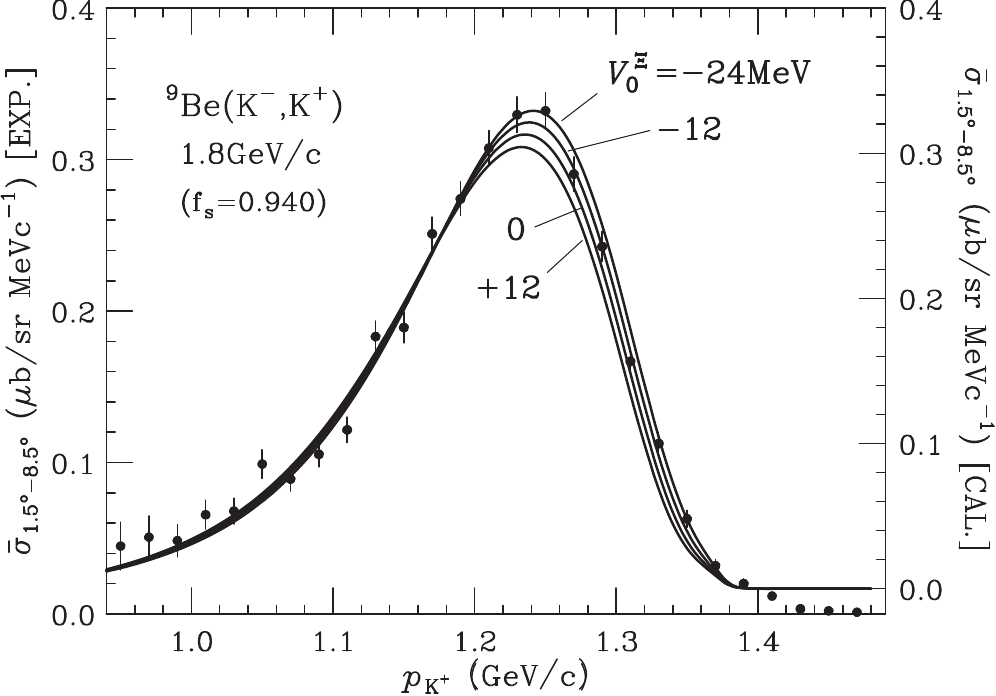} 
\caption{WS fit of the BNL-AGS E906 $^9$Be($K^-,K^+$) spectrum. 
Figure adapted from Ref.~\cite{HH21}.} 
\label{fig:9Be} 
\end{figure}

Nuclear configurations of $\Xi$ hyperons are poorly known~\cite{GHM16,HN18}. 
Because of the large momentum transfer in the standard $(K^-,K^+)$ production 
reaction, induced by the two-body $K^-p\to K^ + \Xi^-$ strangeness exchange 
reaction, $\Xi^-$ hyperons are produced dominantly in the quasi-free continuum 
region. No $\Xi^-$ nor $\Lambda\Lambda$ nuclear bound states have ever 
been observed unambiguously in such experiments~\cite{E224,E885,E906}. 
Nevertheless, an attractive $\Xi$-nuclear Woods-Saxon (WS) potential 
of depth $V_{\Xi}=17\pm 6$~MeV~\cite{HH21} was deduced recently from the 
$^9$Be($K^-,K^+$) quasi-free $\Xi^-$ spectrum shape~\cite{E906} shown in 
Fig.~\ref{fig:9Be}. 

A potential depth $V_{\Xi}=17\pm 6$~MeV is considerably larger than 
$V_{\Xi}\lesssim 10$~MeV deduced from strong-interaction models, e.g., 
HALQCD~\cite{HALQCD20} (confirmed in $p\Xi^-$ femtoscopy~\cite{ALICE19}), 
EFT@NLO~\cite{HM19,K19} and RMF~\cite{Gaitanos21}. A notable exception is 
provided by versions ESC16*(A,B) of the latest Nijmegen model in which values 
of $V_{\Xi}$ higher than 20~MeV arise~\cite{ESC16}. However, these $V_{\Xi}$ 
values get reduced by substantial $\Xi NN$ three-body contributions within 
the same ESC16* model. 

Here we focus on $\Xi$ nuclear constraints derived by observing $\Xi^-$ 
capture events in exposures of light-emulsion CNO nuclei to the $(K^-,K^+)$ 
reaction. A small fraction of the produced high-energy $\Xi^-$ hyperons 
slows down in the emulsion, undergoing an Auger process to form high-$n$ 
atomic states, and cascades down radiatively. Strong-interaction capture 
takes over atomic radiative cascade in a $3D$ atomic orbit bound by 126, 175, 
231~keV in C, N, O, respectively, affected to less than 1~keV by the strong 
interaction~\cite{BFG99}. Nevertheless, captures from a lower orbit have also 
been observed, as follows. 

\begin{table}[!h] 
\centering
\caption{Twin-$\Lambda$ two-body $\Xi^-$ capture events from KEK and J-PARC 
emulsion work. Deduced $\Xi^-$-nuclear binding energies $B_{\Xi^-}$ are 
contrasted with purely Coulomb $2P$ atomic binding energies $B_{\Xi^-}^{2P}$.} 
\begin{tabular}{cccccc} 
\hline
Experiment & Event & $^{A}Z$ & ${_{\Lambda}^{A^{\prime}}}{Z^{\prime}} + 
{_{\Lambda}^{A^{\prime\prime}}}{Z^{\prime\prime}}$ & $B_{\Xi^-}$ (MeV) & 
$B_{\Xi^-}^{2P}$ (MeV)  \\
\hline
KEK E176~\cite{E176} & 10-09-06 & $^{12}$C & \lamb{4}{H}+\lamb{9}{Be} 
& 0.82$\pm$0.17 & 0.285 \\
KEK E176~\cite{E176} & 13-11-14 & $^{12}$C & \lamb{4}{H}+\lamb{9}{Be}$^{\ast}$ 
& 0.82$\pm$0.14 & 0.285 \\
KEK E176~\cite{E176} & 14-03-35 & $^{14}$N & \lamb{3}{H}+\lam{12}{B} 
& 1.18$\pm$0.22 & 0.393 \\
KEK E373~\cite{E373b} & KISO & $^{14}$N & \lamb{5}{He}+\lam{10}{Be}$^{\ast}$ 
& 1.03$\pm$0.18 & 0.393 \\
J-PARC E07~\cite{E07a} & IBUKI & $^{14}$N & \lamb{5}{He}+\lam{10}{Be} 
& 1.27$\pm$0.21 & 0.393 \\
\hline 
\end{tabular} 
\label{tab:twin} 
\end{table} 

Listed in Table~\ref{tab:twin} are {\it all} two-body $\Xi^-$ 
capture events $\Xi^- + {^{A}Z} \to {_{\Lambda}^{A^{\prime}}}{Z^{\prime}} 
+ {_{\Lambda}^{A^{\prime\prime}}}{Z^{\prime\prime}}$ to twin single-$\Lambda$ 
hypernuclei uniquely identified in KEK and J-PARC light-nuclei emulsion 
$K^-$ exposures~\cite{E176,E373b,E07a,E07b}. Expecting $\Lambda$ hyperons 
in $\Xi^-p\to\Lambda\Lambda$ capture to form a spin $S=0\,\,1s_{\Lambda}^2$ 
configuration, the initial $\Xi^-$ hyperon and the proton on which it is 
captured must satisfy $l_{\Xi^-}=l_p$~\cite{Zhu91}, which for $p$-shell 
nuclear targets favors the choice $l_{\Xi^-}=1$. Indeed, all the listed 
events are consistent with $\Xi^-$ capture from Coulomb-assisted $1p_{\Xi^-}$ 
{\it nuclear} states, with $B_{\Xi^-}^{1p}$ larger by about 0.5~MeV than 
the corresponding $2P$ {\it atomic} binding energies $B_{\Xi^-}^{2P}$. 
Not listed in the table are multi-body capture events requiring undetected 
capture products, mostly neutrons, besides a pair of single-$\Lambda$ 
hypernuclei. Two such events~\cite{E07b}, KINKA (KEK-E373) and 
IRRAWADDY (J-PARC E07), correspond to a few MeV $\Xi^-$ binding each, 
suggesting $\Xi^-$ capture from $1s_{\Xi^-}$ nuclear states. Given that 
$1s_{\Xi^-}$ capture rates are of order 1\% of the $1p_{\Xi^-}$ capture 
rates~\cite{Zhu91,Koike17}, this poses a problem. Its likely resolution 
is discussed below.

\section{$\Xi$ nuclear optical potential} 
\label{sec:Vopt} 

$\Xi^-$ atomic and nuclear bound states in $N=Z$ nuclei such as $^{12}$C and 
$^{14}$N are calculated using a finite-size Coulomb potential $V_c^{\Xi^-}$, 
including vacuum-polarization terms, plus a `$t\rho$' optical potential 
$V_{\rm opt}^{\Xi}$~\cite{FG21} where $t$ is a spin-isospin averaged in-medium 
$\Xi N$ $t$-matrix and $\rho=\rho_n+\rho_p$ is a nuclear density normalized to 
the number of nucleons $A$. For $V_{\rm opt}^{\Xi}$ we adopt a form applied in 
Ref.~\cite{FG23a,FG23b} to $V_{\rm opt}^{\Lambda}$:   
\begin{equation} 
V_{\rm opt}^{\Xi}(r)=-\frac{2\pi}{\mu_{\Xi}}\,b_0^A(\rho)\,\rho(r),\,\,\,\,\,\,
\,\,\,b_0^A(\rho)=\frac{{\rm Re}\,b_0^A}{1+(3k_F/2\pi){\rm Re}\,b_0^A}+
{\rm Im}\,b_0. 
\label{eq:Vopt} 
\end{equation}
Here $\mu_{\Xi}$ is the $\Xi^-$-nucleus reduced mass, $b_0^A(\rho)$ is an 
effective density-dependent $\Xi N$ isoscalar c.m. scattering amplitude, 
$b_0^A=(1+\frac{A-1}{A}\frac{\mu_{\Xi}}{m_N})b_0$ transforms $b_0$ from 
the $\Xi N$ c.m. frame to the $\Xi$-nucleus c.m. frame and $k_F$ is the 
Fermi momentum associated with density $\rho$, $k_F^3=3{\pi}^2\rho/2$. 
This form of $V_{\rm opt}^{\Xi}$ accounts for long-range Pauli correlations 
in $\Xi N$ in-medium multiple scatterings, starting at $\rho^{4/3}$ when 
$b_0^A(\rho)$ is expanded in powers of the density $\rho$~\cite{DHL71,WRW97}. 
Shorter-range correlation terms, arising in the present context from 
three-body $\Xi NN$ interactions, start at $\rho^2$ and are briefly 
discussed in the concluding section. 

For $N=Z$ nuclear densities we assumed $\rho_n=\rho_p$ and identified 
the r.m.s. radius of $\rho_p$ with that of the nuclear charge density. 
Folding reasonably chosen $\Xi N$ interaction ranges other than corresponding 
to the proton charge radius, varying the spatial form of the charge density, 
or introducing realistic differences between neutron and proton r.m.s. radii, 
made little difference: in $^{12}$C, for example, all such calculated binding 
energies varied within 20\% of the $\pm$0.15~MeV measurement uncertainty of 
$B_{\Xi^-}^{1p}(^{12}$C) in Table~\ref{tab:twin}. 

For a given absorptivity of Im$\,b_0=0.01$~fm in Eq.~(\ref{eq:Vopt}), $B_{
\Xi^-}^{1p}(^{12}$C)=0.82$\pm$0.15~MeV was fitted by Re$\,b_0=0.495\pm 0.030
$~fm which in the limit $A\to\infty$ and $\rho(r)\to\rho_0 = 0.17$~fm$^{-3}$ 
leads to a depth value $V_{\Xi}=21.2\pm 0.7$~MeV in nuclear matter. This value 
is compatible with that derived from AGS-E906 as shown in Fig.~\ref{fig:9Be} 
and is in agreement with values 21--24~MeV extracted from old emulsion 
events~\cite{DG83}. Disregarding Pauli correlations by setting $k_F = 0$ 
leads to almost 15\% increase of the depth, whereas doubling Im$\,b_0$ 
increases the fitted Re$\,b_0$ by only 1\%~\cite{FG21}.

\section{$1s_{\Xi^-}$ states in $^{14}$N?} 
\label{sec:KINKA} 

\begin{figure}[!h] 
\centering 
%\sidecaption 
\includegraphics[width=0.7\textwidth]{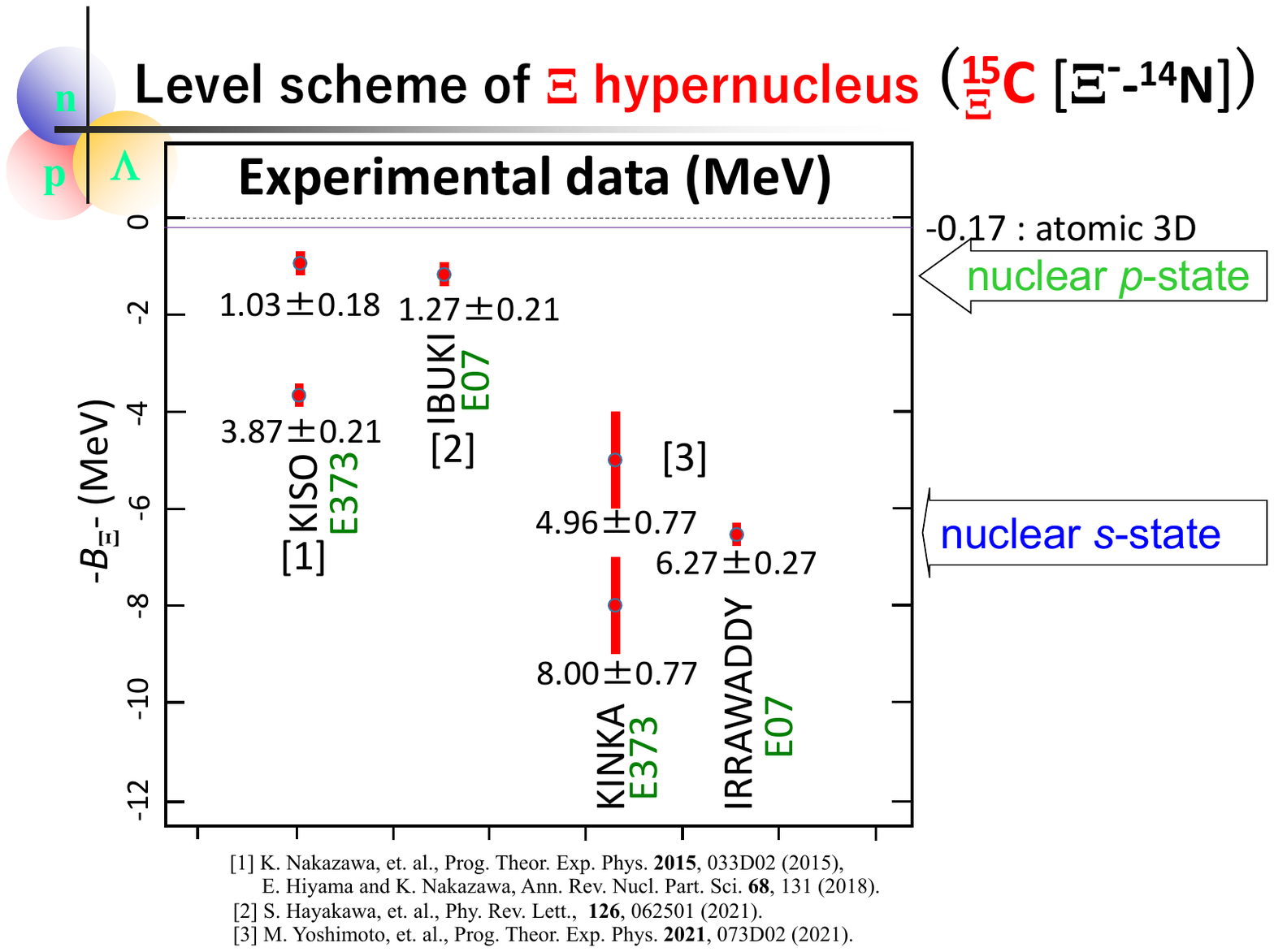} 
\caption{Binding energies of $\Xi^-$ nuclear states in $^{14}$N deduced from 
$\Xi^-$ capture events identified by their twin-$\Lambda$ hypernuclear decays 
in KEK-E373 and J-PARC E07 emulsion experiments. Figure provided by Dr. Kazuma 
Nakazawa, based on recent results from Refs.~\cite{E373b,E07a,E07b}.} 
\label{fig:nakazawa} 
\end{figure} 

In addition to the KISO and IBUKI $\Xi^-_{1p}$ twin-$\Lambda$ capture events 
listed in Table~\ref{tab:twin}, two new $^{14}$N twin-$\Lambda$ capture events 
were reported recently~\cite{E07b}, KINKA from KEK E373 and IRRAWADDY from 
J-PARC E07, both assigned as $\Xi^-_{1s}$ in Fig.~\ref{fig:nakazawa}. We note 
that $2P\to 1S$ radiative decay rates are of order 1\% of $3D\to 2P$ radiative 
decay rates~\cite{Zhu91,Koike17} suggesting that $\Xi^-$ capture from 
a nuclear $\Xi^-_{1s}$--$^{14}$N state is suppressed to this order 
relative to capture from a nuclear $\Xi^-_{1p}$--$^{14}$N state. 
Assigning a $\Xi^-_{1s}$--$^{14}$N bound state to IRRAWADDY or to 
KINKA is therefore questionable. 

\begin{figure}[!h] 
\centering  
%\sidecaption 
\includegraphics[width=0.7\textwidth]{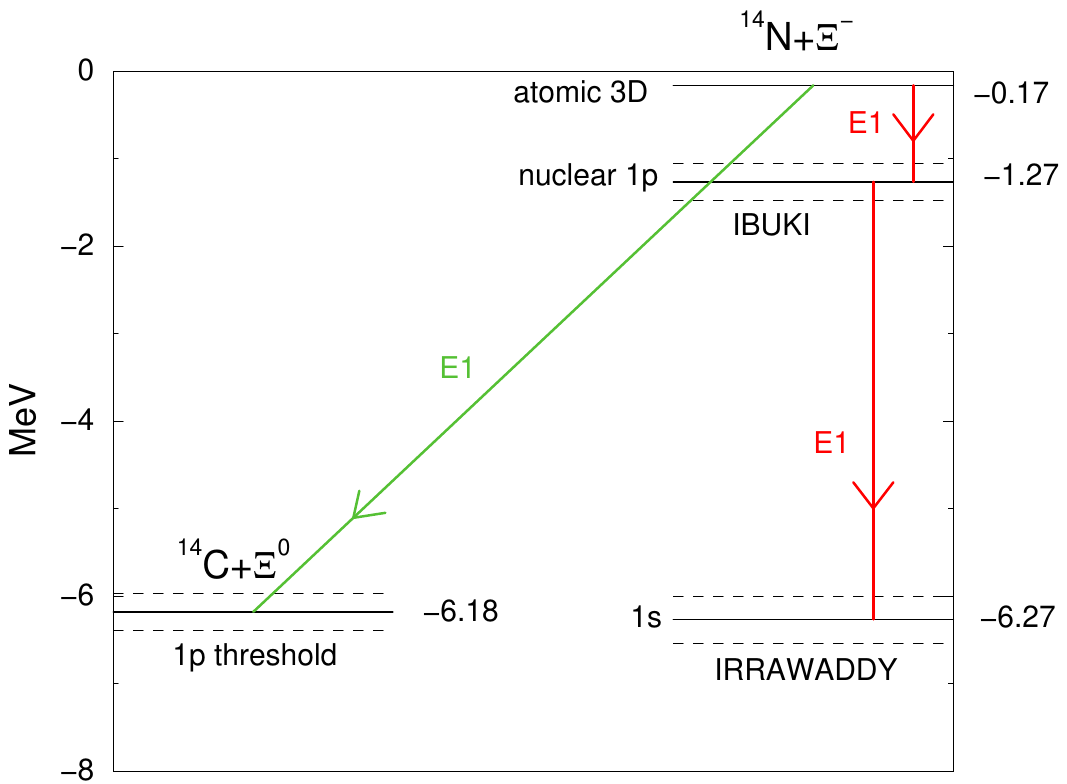} 
\caption{Level diagram of $\Xi^-$--$^{14}$N. Shown on the right are 
a $3D$ atomic state and $1p,1s$ nuclear states assigned, respectively, 
to E07 $\Xi^-$ capture events IBUKI~\cite{E07a} and IRRAWADDY~\cite{E07b}. 
The $\Xi^0$+$^{14}$C threshold at $-$6.18~MeV is marked on the left. 
Electromagnetic $E1$ transitions deexciting the $\Xi^-_{3D}$ atomic state to 
lower, slightly mixed together $\Xi^-$--$^{14}$N and $\Xi^0_{1p}$--$^{14}$C 
nuclear states, are marked by red and green arrowed lines, respectively. 
A near-threshold $\Xi^0_{1p}$--$^{14}$C state on the left provides an 
alternative interpretation of IRRAWADDY~\cite{FG23}.} 
\label{fig:FG23} 
\end{figure} 

It has been suggested by us recently~\cite{FG23} that IRRAWADDY is a 
near-threshold $\Xi^0_{1p}$--$^{14}$C bound state that has nothing to do with 
a $\Xi^-_{1s}$--$^{14}$N bound state claimed by E07. The Coulomb potential's 
role in forming a {\it Coulomb assisted} $\Xi^-_{1p}$--$^{14}$N nuclear 
bound state is replaced for $\Xi^0_{1p}$--$^{14}$C($T=1$) by adding 
a strong-interaction Lane term for isospin $T\neq 0$ nuclear cores in 
$V_{\rm opt}$ of Eq.~(\ref{eq:Vopt}). The sign and strength of this Lane term 
relative to Re~$b_0$ were estimated in Ref.~\cite{FG23} from the sign and 
strength of $V_{\tau}$ relative to $V_0$ in the $\Xi N$ $s$-wave HALQCD 
underlying interaction~\cite{HALQCD20} 
\begin{equation} 
V_{\Xi N}=V_0+V_{\sigma}{\vec\sigma}_{\Xi}\cdot{\vec\sigma}_N+V_{\tau}
{\vec\tau}_{\Xi}\cdot{\vec\tau}_N+V_{\sigma\tau}{\vec\sigma}_{\Xi}\cdot
{\vec\sigma}_N\,{\vec\tau}_{\Xi}\cdot{\vec\tau}_N \, , 
\label{eq:V2body} 
\end{equation}
and were found sufficient to bind $\Xi^0_{1p}$--$^{14}$C near threshold, 
within the J-PARC E07 experimental uncertainty of IRRAWADDY, as shown in 
Fig.~\ref{fig:FG23}. Here, the introduction of $V_{\sigma\tau}$ causes the 
newly considered $\Xi^0_{1p}$--$^{14}$C state to get slightly mixed with 
IBUKI's $\Xi^-_{1p}$--$^{14}$N bound state, sufficiently to make the 
$\sim$6~MeV $E1$ radiative deexcitation of the $\Xi^-_{3D}$--$^{14}$N 
atomic state to the dominantly $\Xi^0_{1p}$--$^{14}$C nuclear state as 
strong as to the $\sim$1~MeV deexcitation to the IBUKI $\Xi^-_{1p}-{^{14}}$N 
nuclear state. Assigning a $\Xi^0_{1p}$--$^{14}$C bound state structure 
to IRRAWADDY contrasts with viewing it as a $\Xi^-_{1s}$--$^{14}$N state 
motivated largely by IRRAWADDY's binding energy of a few MeV.

It is worth recalling that in spite of limiting discussion to the fairly 
narrow IRRAWADDY, given that a KINKA+IRRAWADDY weighted average of 
$B_{\Xi^-}$=6.13$\pm$0.25~MeV or 6.46$\pm$0.25~MeV differs little from 
IRRAWADDY's own value of $B_{\Xi^-}$=6.27$\pm$0.27~MeV, our arguments 
apply equally well to either one of KINKA's considerably broader versions.

\section{Discussion}
\label{sec:disc}

\begin{table}[htb] 
\centering 
\caption{Input (underlined) and calculated mean values of $\Xi^-$ binding 
energies in optical potential fits, plus resulting $\Xi N$ and $\Xi NN$ 
induced $\Xi$ nuclear potential depths $V^{(2)}_{\Xi}$ and $V^{(3)}_{\Xi}$, 
respectively, at nuclear-matter density $\rho_0=0.17\,$fm$^{-3}$, and their 
correlated sum $V_{\Xi}$; see text. Potentials and energies are given in MeV.} 
\begin{tabular}{ccccccc} 
\hline 
$B_{\Xi^-}^{1p}(^{12}$C) & $B_{\Xi^-}^{1s}(^{14}$N) & $B_{\Xi^-}^{1p}(^{14}$N) 
& $V_{\Xi}^{(2)}$ & $V_{\Xi}^{(3)}$ & $V_{\Xi}$ & $B_{\Xi^-}^{1s}(^{11}$B) \\
\hline
$\underline{0.82}$ & 11.79 & 1.94 & 21.2$\pm$0.7&--&21.2$\pm$0.7 & 9.00 \\
0.32 & $\underline{6.27}$ & 0.52 & 13.6$\pm$0.4&--&13.6$\pm$0.4 & 4.20 \\ 
$\underline{0.82}$ & $\underline{8.00}$ & $\underline{1.27}$ & 26.4$\pm$2.6 & 
$-$15.4$\pm$5.7 & 11.0$\pm$3.1 & 6.29 \\ 
$\underline{0.82}$ & $\underline{6.27}$ & $\underline{1.27}$ & 30.6$\pm$1.7 & 
$-$28.2$\pm$3.9 & 2.4$\pm$2.2 & 5.15 \\ 
\hline 
\end{tabular} 
\label{tab:targets} 
\end{table} 

Some $\Xi$-nuclear scenarios are outlined in Table~\ref{tab:targets}. 
Choosing $B_{\Xi^-}^{1p}$=0.82$\pm$0.15~MeV for the two KEK-E176 $^{12}$C 
events listed in Table~\ref{tab:twin} to fit the strength $b_0$ of the $\Xi N$ 
induced $\Xi$-nuclear attractive optical potential $V_{\rm opt}^{\Xi}$ in 
Eq.~(\ref{eq:Vopt}) results in several other $\Xi^-$ nuclear binding energies 
listed in the first row. Choosing instead the J-PARC E07 $^{14}$N IRRAWADDY 
event~\cite{E07b} with $B_{\Xi^-}^{1s}$=6.27$\pm$0.27~MeV as input results 
in values listed in the second row of the table. Clearly, these two sets 
of results differ strongly for the $\Xi^-_{1s}$--$^{11}$B binding energies 
discussed below and for the $\Xi$-nuclear attractive potential depths 
$V_{\Xi}^{(2)}$. As for the Coulomb assisted $\Xi^-_{1p}$--$^{12}$C nuclear 
state, we note that it lies deeper by merely 40~keV than the Coulomb $2P$ 
atomic state in $^{12}$C ($B_{\Xi^-}^{2P}=283$~keV) when constrained by 
IRRAWADDY in the second row, strongly disagreeing with the 540$\pm$150~keV 
extra strong-interaction binding deduced from the KEK E176 events~\cite{E176} 
underlined in the first row. 

The next two rows in Table~\ref{tab:targets} report on fitting {\it two} 
$\Xi$-nucleus interaction parameters, $b_0$ for the $\Xi N$ induced attractive 
$V_{\rm opt}^{\Xi}$, Eq.~(\ref{eq:Vopt}), and $B_0$ for a $\Xi NN$ induced 
potential term 
\begin{equation} 
\delta V^{\Xi}_{\rm opt}(r)=\frac{2\pi}{\mu_{\Xi}}\,(1+\frac{A-2}{A}
\frac{\mu_{\Xi}}{2m_N})\,B_0\,\frac{\rho^2(r)}{\rho_0}  
\label{eq:V3body} 
\end{equation}
introduced in our recent work on the content of the $\Lambda$-nuclear optical 
potential~\cite{FG23a,FG23b}. Both $^{12}$C and $^{14}$N $\Xi^-_{1p}$ bound 
states are used for input, along with KINKA's higher-binding option in the 
third row or IRRAWADDY in the fourth row for $B_{\Xi^-}^{1s}(^{14}$N) to allow 
for some variation. Both fits are acceptable, $\chi^2 < 1$, with a substantial 
$\Xi NN$ induced repulsive $\delta V^{\Xi}_{\rm opt}$ almost doubling its 
strength as $B_{\Xi^-}^{1s}(^{14}$N) input is decreased from 8.00$\pm$0.77~MeV 
in the third row to 6.27$\pm$0.27~MeV in the fourth row. We note that 
the $\Xi N$ induced potential $V_{\Xi}^{(2)}$ depth values obtained when 
$\delta V^{\Xi}_{\rm opt}$ is introduced increase farther away from the 
considerably smaller values of $V_{\Xi}^{(2)}$ obtained in recent theoretical 
models~\cite{HALQCD20,HM19}, while the total depth $V_{\Xi}$ decreases farther 
away from $V_{\Xi}=17\pm 6$~MeV suggested by the $^9$Be($K^-,K^+$) spectrum in 
Fig.~\ref{fig:9Be}. 

The solution proposed here to the difficulty of interpreting IRRAWADDY as
a $\Xi^-_{1s}$ bound state in $^{14}$N is by pointing out that it could
correspond to a $\Xi^0_{1p}$--$^{14}$C bound state, something that cannot
occur kinematically in the other light-emulsion nuclei $^{12}$C and $^{16}$O. 
Given that in this nuclear mass range capture rates from $1s_{\Xi^-}$ states 
are estimated to be two orders of magnitude below capture rates from $1p_{
\Xi^-}$ states~\cite{Zhu91,Koike17}, our $\Xi^0_{1p}$--$^{14}$C assignment 
addresses satisfactorily the capture rate hierarchy. 

Regarding $1s_{\Xi^-}$ states, J-PARC E70 $^{12}$C$(K^-,K^+)^{12}_{~\Xi}$Be 
experiment with record 2~MeV FWHM simulated resolution~\cite{Go22}, following 
an earlier experiment E05 with 5.4~MeV FWHM resolution~\cite{Go20}, aims 
particularly to observe $\Xi^-_{1s}-{^{11}}$B signals. The last column in 
Table~\ref{tab:targets} lists a wide range of predicted $B_{\Xi^-}^{1s}(^{11}
$B) values depending on which $\Xi^-$-capture data are accepted. Corrections 
of order 0.5 MeV are likely from the three spin-isospin $\Xi N$ terms in 
Eq.~(\ref{eq:V2body}).

\section*{Acknowledgments} 
\begin{acknowledgement} 
One of us (A.G.) thanks the organizers of MESON2023 in Krakow for support 
and hospitality. This work is part of a project funded by the EU Horizon 
2020 Research \& Innovation Programme under grant agreement 824093. 
\end{acknowledgement}

\end{document}